\input amstex
\TagsOnRight

\font\larger=cmr10 scaled \magstep1
\font\largerr=cmr10 scaled \magstep2
\baselineskip 14 pt
\parskip 5 pt

\noindent {{\bf {\largerr Inverse Scattering Transform and the Theory of Solitons}}}

\vskip 15 pt
\noindent {{\larger TUNCAY AKTOSUN$^{\text {ab}}$}

\vskip -7 pt
\noindent {$^{\text a}$University of Texas at Arlington, Arlington, Texas, USA}
\vskip -7 pt
\noindent {$^{\text b}$Supported in part
by the National Science Foundation under grant DMS-0610494}

\vskip 15 pt
\noindent {{\larger  Article Outline}}

\noindent {Glossary}
\vskip -7 pt

\noindent {I. Definition of the Subject and Its Importance}
\vskip -7 pt

\noindent {II. Introduction}
\vskip -7 pt

\noindent {III. Inverse Scattering Transform (IST)}

\vskip -7 pt

\noindent {IV. The Lax Method}
\vskip -7 pt

\noindent {V. The AKNS Method}
\vskip -7 pt

\noindent {VI. Direct Scattering Problem}
\vskip -7 pt

\noindent {VII. Time Evolution of the Scattering Data}
\vskip -7 pt

\noindent {VIII. Inverse Scattering Problem}
\vskip -7 pt

\noindent {IX. Solitons}

\vskip -7 pt

\noindent {X. Future Directions}
\vskip -7 pt

\noindent {XI. Bibliography}

\vskip 15 pt

\noindent {{\bf {\larger Glossary}}

\noindent {\bf AKNS method} A method introduced by Ablowitz,
Kaup, Newell, and Segur in 1973
that identifies the nonlinear partial differential equation (NPDE)
associated with a given first-order system
of linear ordinary differential equations (LODEs) so that
the initial value problem (IVP) for that NPDE can be solved by the
inverse scattering transform (IST) method.

\noindent {\bf Direct scattering problem} The problem of determining
the scattering data corresponding to a given potential
in a differential equation.

\noindent {\bf Integrability} A NPDE is said to be integrable if
its IVP can be solved via an IST.

\noindent {\bf Inverse scattering problem} The problem of determining
the potential that corresponds to a given set of scattering data
in a differential equation.

\noindent {\bf Inverse scattering transform}
A method introduced in 1967 by
Gardner, Greene, Kruskal, and Miura that yields a solution to
the IVP for a NPDE with the help of the solutions to the direct
and inverse scattering problems for an associated
LODE.

\noindent {\bf Lax method} A method introduced by Lax in 1968
that determines the integrable NPDE associated with a given LODE so that
the IVP for that NPDE can be solved with the help of an IST.

\noindent {\bf Scattering data} The scattering data associated with a LODE
usually consists of a reflection coefficient
which is a function of the spectral parameter $\lambda,$ a finite
number of constants $\lambda_j$ that correspond to the poles of the
transmission coefficient in the upper half complex plane, and
the bound-state norming constants whose number for each
bound-state pole $\lambda_j$ is the same as the order of that
pole. It is desirable that the potential
in the LODE is uniquely determined by the corresponding scattering data
and vice versa.

\noindent {\bf Soliton}
The part of a solution to
an integrable NPDE due to a pole of the transmission coefficient
in the upper half complex plane.
The
term soliton was introduced by Zabusky and Kruskal in 1965
to denote a solitary wave pulse with a particle-like
behavior in the solution to the Korteweg-de Vries (KdV) equation.

\noindent {\bf Time evolution of the scattering data} The evolvement
of the scattering data from its initial value $S(\lambda,0)$ at $t=0$ to
its value $S(\lambda,t)$ at a later time $t.$

\vskip 20 pt

\noindent {{\bf {\larger I. Definition of the Subject and Its Importance}}}

A general theory
to solve NPDEs does not seem to exist. However, there are certain
NPDEs, usually first order in time, for which the corresponding
IVPs can be solved by the IST method.
Such NPDEs are sometimes referred to as
integrable evolution equations. Some exact solutions
to such equations may be available in terms of
elementary functions, and such solutions
are important to understand nonlinearity better and they
may also be useful in testing accuracy of numerical
methods to solve such NPDEs.

Certain special
solutions to some of such NPDEs exhibit particle-like behaviors. A
single-soliton solution is usually a localized
disturbance that retains its shape but only changes its location in time.
A multi-soliton solution consists of several solitons that
interact nonlinearly when they are close to each
other but come out of such interactions unchanged in shape except for a
phase shift.

Integrable NPDEs have important physical applications.
For example,
the KdV equation is used to describe [14,23]
surface water waves in long, narrow, shallow canals; it also arises [23]
in the description of hydromagnetic waves in a cold plasma,
and ion-acoustic waves in anharmonic crystals.
The nonlinear Schr\"odinger (NLS) equation arises in modeling [24]
electromagnetic waves in optical
fibers as well as surface waves in deep waters. The sine-Gordon equation
is helpful [1] in analyzing the magnetic field in a
Josephson junction (gap between two superconductors).

\vskip 15 pt
\noindent {{\bf {\larger II. Introduction}}}
\vskip 3 pt

The first observation of a soliton
was made in 1834 by the Scottish engineer John Scott Russell
at the
Union Canal between Edinburgh and Glasgow.
Russell reported [21] his observation
to the British Association of the Advancement of
Science in September 1844, but he did not seem
to be successful in convincing
the scientific community. For example, his contemporary George Airy,
the influential
mathematician of the time, did not believe in the existence of
solitary water waves [1].

The Dutch mathematician Korteweg and his doctoral student
de Vries published [14] a paper in 1895 based on
de Vries' Ph.D. dissertation, in which surface
waves in shallow, narrow canals were modeled
by what is now known as the KdV equation.
The importance of this paper was not understood until 1965 even though
it contained as a special solution what is now known as the one-soliton
solution.

Enrico Fermi in his summer visits to the Los Alamos
National Laboratory, together with J. Pasta and S. Ulam,
used the computer named Maniac I to computationally
analyze a one-dimensional dynamical system
of 64 particles in which adjacent particles
were joined by springs where the forces also
included some nonlinear terms. Their main
goal was to determine the rate of approach to
the equipartition of energy among
different modes of the system. Contrary
to their expectations there was little tendency towards
the equipartition of energy but instead the almost
ongoing recurrence to the initial state,
which was puzzling. After Fermi died in November 1954,
Pasta and Ulam completed their last few computational
examples and finished writing a preprint [11],
which was never published as a journal article.
This preprint appears in Fermi's Collected Papers [10]
and is also available on the internet [25].

In 1965 Zabusky and Kruskal explained [23] the Fermi-Pasta-Ulam
puzzle in terms of solitary wave solutions to the
KdV equation. In their numerical analysis they observed
``solitary-wave pulses," named such pulses
``solitons" because of their particle-like behavior,
and noted that such pulses interact with each other nonlinearly
but come out of interactions unaffected in size or shape except
for some phase shifts. Such unusual
interactions among solitons generated a lot of
excitement, but at that time
no one knew how to solve the IVP for
the KdV equation, except numerically.
In 1967 Gardner, Greene, Kruskal, and Miura presented
[12] a method, now known as the IST, to solve that IVP, assuming that the initial
profile $u(x,0)$ decays to zero sufficiently rapidly
as $x\to\pm\infty.$ They showed that the integrable NPDE,
i.e. the KdV equation,
$$u_t-
6uu_x+
u_{xxx}=0,\tag 2.1$$
is associated with a LODE, i.e.
the 1-D Schr\"odinger equation
$$-\displaystyle\frac{d^2\psi}{dx^2}+u(x,t)\,\psi=
k^2\psi,\tag 2.2$$
and that the solution
$u(x,t)$ to (2.1) can be recovered from the initial
profile $u(x,0)$ as explained in the diagram given in Section III.
They also explained that soliton solutions to the KdV equation
correspond to a zero reflection coefficient
in the associated scattering data. Note that the subscripts
$x$ and $t$ in (2.1) and throughout denote the partial derivatives
with respect to those variables.

In 1972 Zakharov and Shabat showed [24] that the IST method
is applicable also to
the IVP for the NLS equation
$$iu_t+u_{xx}+2|u|^2u=0,\tag 2.3$$
where $i$ denotes the imaginary number $\sqrt{-1}.$
They proved that the associated LODE is
the first-order linear system
$$\cases \displaystyle\frac{d\xi}{dx}=-i\lambda\xi+u(x,t)\,\eta,\\
\noalign{\medskip}
\displaystyle\frac{d\eta}{dx}=i\lambda \eta-\overline{u(x,t)}\,\xi,\endcases
\tag 2.4$$
where $\lambda$ is the spectral parameter and
an overline denotes complex conjugation.
The system (2.4) is now known as the
Zakharov-Shabat system.

Soon afterwards, again in 1972 Wadati
showed in a one-page
publication [22] that
the IVP for the modified Korteweg-de Vries (mKdV) equation
$$u_t+6u^2u_x+u_{xxx}=0,\tag 2.5$$
can be solved with the help of the inverse scattering
problem for the linear system
$$\cases \displaystyle\frac{d\xi}{dx}=-i\lambda\xi+u(x,t)\,\eta,\\
\noalign{\medskip}
\displaystyle\frac{d\eta}{dx}=i\lambda \eta-u(x,t)\,\xi.\endcases
\tag 2.6$$

Next, in 1973
Ablowitz, Kaup, Newell, and Segur showed [2,3] that the IVP for
the sine-Gordon equation
$$u_{xt}=\sin u,$$
can be solved in the same way
by exploiting the inverse scattering problem associated with the
linear system
$$\cases \displaystyle\frac{d\xi}{dx}=-i\lambda\xi-\displaystyle\frac12\,u_x(x,t)\,\eta,\\
\noalign{\medskip}
\displaystyle\frac{d\eta}{dx}=i\lambda \eta+\displaystyle\frac12\,u_x(x,t)\,\xi.\endcases$$
Since then, many other NPDEs have been discovered
to be solvable by the IST method.

Our review is organized as follows.
In the next section we explain the idea behind the IST.
Given a LODE known to be associated with an integrable NPDE,
there are two primary methods
enabling us to determine the corresponding NPDE. We review those two methods,
the Lax method and the AKNS method,
in Section IV and in Section V, respectively.
In Section VI we introduce
the scattering data associated with a LODE
containing a spectral parameter and a
potential, and we illustrate it
for the Schr\"odinger equation and for the Zakharov-Shabat
system. In Section VII we explain the time evolution
of the scattering data and indicate how the scattering data sets
evolve for
those two LODEs.
In Section VIII we summarize the Marchenko
method to solve
the inverse scattering problem for
the Schr\"odinger equation and that for the Zakharov-Shabat
system, and we outline how the solutions to the IVPs for
the KdV equation and the NLS
equation are obtained with the help of the IST.
In Section IX we present soliton solutions to the KdV and NLS equations.
A brief conclusion is provided in Section X.

\vskip 15 pt
\noindent {{\bf {\larger III. Inverse Scattering Transform}}}
\vskip 3 pt

\noindent Certain NPDEs are classified as integrable in the
sense that their corresponding IVPs can be solved
with the help of an IST.
The idea behind the IST method is as follows: Each integrable NPDE is associated with
a LODE (or a system of LODEs) containing
a parameter $\lambda$ (usually known as the spectral parameter), and
the solution $u(x,t)$ to the NPDE appears as a coefficient
(usually known as the potential) in the corresponding LODE.
In the NPDE the quantities $x$ and $t$ appear as independent variables
(usually known as the spatial
and temporal coordinates, respectively), and in the LODE
$x$ is an independent variable and $\lambda$ and $t$ appear as parameters.
It is usually the case that $u(x,t)$ vanishes at each fixed $t$ as
$x$ becomes infinite
so that a scattering scenario can be created for the related
LODE, in which the potential $u(x,t)$ can uniquely be associated
with some scattering data $S(\lambda,t).$
The problem of determining $S(\lambda,t)$ for all
$\lambda$ values from $u(x,t)$ given for all $x$ values
is known as the direct scattering problem
for the LODE. On the other hand, the problem of determining $u(x,t)$ from
$S(\lambda,t)$ is known as the inverse scattering problem for that
LODE.

The IST method for an integrable NPDE can be explained
with the help of the diagram

\vskip 12 pt

\qquad\qquad\qquad\qquad
{$\CD u(x,0) @>\text{direct scattering for LODE at } t=0>> S(\lambda,0) \\
@V{\text{solution to NPDE}}VV @VV{   \text{time evolution of scattering data}  }V
 \\
u(x,t) @<<\text{inverse scattering for LODE at time } t< S(\lambda,t)
\endCD$}

\vskip 8 pt

\noindent In order to solve the IVP for the NPDE, i.e. in
order to determine $u(x,t)$ from $u(x,0),$ one needs to perform the following three steps:

\item{(i)} Solve the corresponding direct scattering problem
for the associated LODE at $t=0,$ i.e. determine the initial scattering data
$S(\lambda,0)$ from the initial potential $u(x,0).$

\item{(ii)} Time evolve the scattering data from its initial value
$S(\lambda,0)$ to its value $S(\lambda,t)$ at time $t.$
Such an evolution is usually a simple one and
is particular to each integrable NPDE.

\item{(iii)} Solve the corresponding inverse scattering problem for the associated
LODE at fixed $t,$ i.e. determine
the potential $u(x,t)$ from the scattering data $S(\lambda,t).$

\noindent It is amazing that the resulting $u(x,t)$ satisfies the integrable
NPDE and that the limiting value of $u(x,t)$ as $t\to 0$ agrees with the initial
profile $u(x,0).$

\vskip 10 pt
\noindent {{\bf {\larger IV. The Lax Method}}}
\vskip 3 pt

In 1968 Peter Lax
introduced [15] a method yielding
an integrable NPDE corresponding to a given LODE.
The basic idea behind the Lax method is the following.
Given a linear differential operator $\Cal L$
appearing in the spectral problem $\Cal L\psi=\lambda \psi,$
find an operator $\Cal A$ (the operators
$\Cal A$ and $\Cal L$ are said to form a Lax pair)
such that:
\item{(i)} The spectral parameter
$\lambda$ does not change in time, i.e. $\lambda_t=0.$
\item{(ii)} The quantity $\psi_t-\Cal A\psi$ remains
a solution to the same linear problem $\Cal L\psi=\lambda \psi.$
\item{(iii)} The quantity
$\Cal L_t+\Cal L \Cal A-\Cal A\Cal L$ is a multiplication operator,
i.e. it is not a differential operator.

 From condition (ii) we get
$$\Cal L\left(\psi_t-\Cal A\psi\right)=\lambda\left(\psi_t-\Cal A
\psi\right),\tag 4.1$$
and with the help of
$\Cal L\psi=\lambda \psi$ and $\lambda_t=0,$ from (4.1) we obtain
$$
\Cal L\psi_t-\Cal L \Cal A\psi=
\lambda\psi_t-\Cal A\left(\lambda\psi\right)
=\partial_t\left( \lambda\psi\right)-\Cal A\Cal L \psi
=\partial_t\left(\Cal L\psi\right)-\Cal A\Cal L\psi
=\Cal L_t\psi+\Cal L\psi_t-\Cal A\Cal L\psi,\tag 4.2$$
where $\partial_t$ denotes the partial differential operator
with respect to $t.$
After canceling the term
$\Cal L\psi_t$ on the left and right hand sides of (4.2), we get
$$\left(\Cal L_t+\Cal L \Cal A-\Cal A \Cal L\right)\psi=0,$$
which, because of (iii), yields
$$\Cal L_t+\Cal L \Cal A-\Cal A\Cal L=0.\tag 4.3$$
Note that (4.3) is
an evolution equation containing a first-order
time derivative, and
it is the desired integrable NPDE.
The equation (4.3) is often called a compatibility
condition.

Having outlined the Lax method,
let us now illustrate it to derive
the KdV equation (2.1)
 from the Schr\"odinger equation (2.2).
For this purpose, we write the Schr\"odinger equation as
$\Cal L\psi=\lambda\psi$ with $\lambda:=k^2$ and
$$\Cal L:=-\partial_x^2+u(x,t),\tag 4.4$$
where the notation $:=$ is used to indicate
a definition so that the quantity on the left
should be understood as the quantity on the right hand side.
Given the linear differential operator $\Cal L$ defined
as in (4.4), let us
try to determine the associated
operator $\Cal A$ by assuming that it has the form
$$\Cal A=\alpha_3\partial_x^3+\alpha_2 \partial_x^2+\alpha_1\partial_x+\alpha_0,
\tag 4.5$$
where
the coefficients $\alpha_j$ with $j=0,1,2,3$ may depend
on $x$ and $t,$ but not on the spectral parameter $\lambda.$
Note that $\Cal L_t=u_t.$ Using (4.4) and (4.5) in (4.3),
we obtain
$$(\ )\partial_x^5+(\ )\partial_x^4
+(\ )\partial_x^3+(\ )\partial_x^2+(\ )\partial_x+(\ )=0,\tag 4.6$$
where, because of (iii), each coefficient denoted
by $(\ )$ must vanish.
The coefficient of $\partial_x^5$ vanishes automatically.
Setting the coefficients of $\partial_x^j$ to zero
for $j=4,3,2,1,$ we obtain
$$\alpha_3=c_1,\quad
\alpha_2=c_2,\quad
\alpha_1=c_3-\displaystyle\frac32 c_1u,\quad
\alpha_0=c_4-\displaystyle\frac34c_1u_x-c_2u,$$
with $c_1,$ $c_2,$ $c_3,$ and $c_4$ denoting arbitrary
constants. Choosing $c_1=-4$ and $c_3=0$ in the
last coefficient in (4.6)
and setting that coefficient to zero, we get
the KdV equation (2.1).
Moreover, by letting $c_2=c_4=0,$ we obtain
the operator $\Cal A$ as
$$\Cal A=-4\partial_x^3+6u\partial_x+3u_x.\tag 4.7$$

For the Zakharov-Shabat system (2.4), we proceed in a
similar way. Let us write it as
$\Cal L\psi=\lambda\psi,$ where the linear differential
operator $\Cal L$ is defined via
$$\Cal L:=i\bmatrix
1&0\\
\noalign{\medskip}
0&-1\endbmatrix\partial_x-i\bmatrix 0&u(x,t)\\
\noalign{\medskip}
\overline{u(x,t)}&0\endbmatrix.$$
Then, the operator $\Cal A$ is obtained as
$$\Cal A=2i\bmatrix
1&0\\
\noalign{\medskip}
0&-1\endbmatrix\partial^2_x
-2i\bmatrix 0&u\\
\noalign{\medskip}
\overline{u}&0\endbmatrix\partial_x
-i\bmatrix
-|u|^2&u_x\\
\noalign{\medskip}
\overline{u}_x&|u|^2\endbmatrix,\tag 4.8$$
and the compatibility condition (4.3) gives us
the NLS equation
(2.3).

For the first-order system (2.6), by writing it as
$\Cal L\psi=\lambda\psi,$ where the
linear operator $\Cal L$ is defined by
$$\Cal L:=i\bmatrix
1&0\\
\noalign{\medskip}
0&-1\endbmatrix\partial_x-i\bmatrix 0&u(x,t)\\
\noalign{\medskip}
u(x,t)&0\endbmatrix,$$
we obtain the corresponding operator $\Cal A$ as
$$\Cal A=-4\bmatrix
1&0\\
\noalign{\medskip}
0&1\endbmatrix\partial^3_x
-6\bmatrix u^2&-u_x\\
\noalign{\medskip}
u_x&u^2\endbmatrix\partial_x
-\bmatrix
6uu_x&-3u_{xx}\\
\noalign{\medskip}
3u_{xx}&6uu_x\endbmatrix,$$
and the compatibility condition (4.3) yields
the mKdV equation
(2.5).

\vskip 15 pt
\noindent {{\bf {\larger V. The AKNS Method}}}
\vskip 3 pt

In 1973 Ablowitz, Kaup, Newell, and Segur
introduced [2,3] another method to determine
an integrable NPDE corresponding to a LODE.
This method is now known as
the AKNS method, and the basic idea behind it
is the following. Given a linear operator $\Cal X$ associated
with the first-order system $v_x=\Cal Xv,$
we are interested in finding
an operator $\Cal T$ (the operators $\Cal T$ and $\Cal X$
are said to form an AKNS pair)
such that:
\item{(i)} The spectral parameter
$\lambda$ does not change in time, i.e. $\lambda_t=0.$
\item{(ii)} The quantity $v_t-\Cal Tv$
is also a solution to $v_x=\Cal Xv,$ i.e. we have
$(v_t-\Cal Tv)_x=\Cal X(v_t-\Cal Tv).$
\item{(iii)} The quantity
$\Cal X_t-\Cal T_x+\Cal X\Cal T-\Cal T\Cal X$ is a
(matrix) multiplication operator, i.e.
it is not a differential operator.

 From condition (ii) we get
$$v_{tx}-\Cal T_xv-\Cal Tv_x=\Cal Xv_t-\Cal X\Cal Tv
=(\Cal Xv)_t-\Cal X_tv-\Cal X\Cal Tv
=(v_x)_t-\Cal X_tv-\Cal X\Cal Tv
=v_{xt}-\Cal X_tv-\Cal X\Cal Tv.\tag 5.1$$
Using
$v_{tx}=v_{xt}$ and replacing $\Cal Tv_x$ by
$\Cal T\Cal X v$ on the left side
and equating the left and right hand sides
in (5.1), we obtain
$$(\Cal X_t-\Cal T_x+\Cal X\Cal T-\Cal T\Cal X)v=0,$$
which in turn, because of (iii),
implies
$$\Cal X_t-\Cal T_x+\Cal X\Cal T-\Cal T\Cal X=0.\tag 5.2$$
We can view (5.2) as an integrable
NPDE solvable with the help of the solutions
to the direct and inverse scattering problems for
the linear system $v_x=\Cal X v.$
Like (4.3), the compatibility condition
(5.2) yields a nonlinear evolution equation
containing a first-order time derivative.
Note that $\Cal X$ contains
the spectral parameter $\lambda,$
and hence $\Cal T$ also depends on $\lambda$ as well.
This is in contrast with the Lax method in the sense that
the operator $\Cal A$ does not contain $\lambda.$

Let us illustrate the AKNS method by deriving the KdV equation (2.1) from
the Schr\"odinger equation (2.2).
For this purpose we write the Schr\"odinger equation, by replacing
the spectral parameter $k^2$ with $\lambda,$ as
a first-order linear system $v_x=\Cal Xv,$ where we have defined
$$v:=\bmatrix \psi_x\\
\noalign{\medskip}
\psi\endbmatrix,\quad \Cal X:=\bmatrix 0&u(x,t)-\lambda\\
\noalign{\medskip}
1&0\endbmatrix.$$
Let us look for $\Cal T$ in the form
$$\Cal T=\bmatrix \alpha&\beta\\
\noalign{\medskip}
\rho&\sigma\endbmatrix,$$
where the entries $\alpha,$ $\beta,$ $\rho,$ and $\sigma$ may depend
on $x,$ $t,$ and $\lambda.$
The compatibility condition (5.2)
yields
$$\bmatrix
-\alpha_x-\beta+\rho(u-\lambda)&
u_t-\beta_x+\sigma(u-\lambda)-\alpha(u-\lambda)\\
\noalign{\medskip}
-\rho_x+\alpha-\sigma&
-\sigma_x+\beta-\rho(u-\lambda)\endbmatrix=\bmatrix 0&0\\
\noalign{\medskip}
0&0\endbmatrix.\tag 5.3$$
The $(1,1),$ $(2,1),$ and $(2,2)$-entries in
the matrix equation (5.3) imply
$$\beta=-\alpha_x+(u-\lambda)\rho,\quad
\sigma=\alpha-\rho_x,\quad
\sigma_x=-\alpha_x.\tag 5.4$$
Then from the $(1,2)$-entry in (5.3) we obtain
$$u_t+\displaystyle\frac12\rho_{xxx}-u_x\rho-2\rho_x(u-\lambda)=0.\tag 5.5$$
Assuming a linear dependence of
$\rho$ on the spectral parameter
and hence letting $\rho=\lambda\zeta+\mu$ in (5.5),
we get
$$2\zeta_x\lambda^2+\left(\displaystyle\frac12\zeta_{xxx}-
2\zeta_xu+2\mu_x-u_x\zeta\right)
\lambda
+\left(u_t
+\displaystyle\frac12\mu_{xxx}-2\mu_xu-u_x\mu\right)=0.$$
Equating the coefficients of each power of
$\lambda$ to zero, we have
$$\zeta=c_1,\quad \mu=\displaystyle\frac12 c_1u+c_2,\quad
u_t-\displaystyle\frac32c_1uu_x-c_2u_x+\displaystyle\frac14
c_1u_{xxx}=0,\tag 5.6$$
with $c_1$ and $c_2$ denoting arbitrary constants.
Choosing $c_1=4$ and $c_2=0,$ from (5.6) we obtain
the KdV equation given in (2.1). Moreover, with the help of (5.4) we get
$$\alpha=u_x+c_3,\quad
\beta=-4\lambda^2+2\lambda u+2u^2-u_{xx},
\quad \rho=4\lambda+2u,\quad
\sigma=c_3-u_x,$$
where $c_3$ is an arbitrary constant.
Choosing $c_3=0,$ we find
$$\Cal T=\bmatrix u_x&-4\lambda^2+2\lambda u+2u^2-u_{xx}\\
\noalign{\medskip}
4\lambda+2u&-u_x\endbmatrix.$$

As for the Zakharov-Shabat system (2.4), writing it
as
$v_x=\Cal X v,$ where we have defined
$$\Cal X:=\bmatrix
-i\lambda&u(x,t)\\
\noalign{\medskip}
-\overline{u(x,t)}&i\lambda\endbmatrix,$$
we obtain the matrix operator $\Cal T$ as
$$\Cal T=\bmatrix
-2i\lambda^2+i|u|^2&2\lambda u+iu_x\\
\noalign{\medskip}
-2\lambda\overline{u}+i\overline{u}_x&
2i\lambda^2-i|u|^2\endbmatrix,$$
and the compatibility condition (5.2) yields
the NLS equation (2.3).

As for the first-order linear system (2.6), by writing it as
$v_x=\Cal X v,$ where
$$\Cal X:=\bmatrix
-i\lambda&u(x,t)\\
\noalign{\medskip}
-u(x,t)&i\lambda\endbmatrix,$$
we obtain the matrix operator $\Cal T$ as
$$\Cal T=\bmatrix
-4i\lambda^3+2i\lambda u^2&4\lambda^2u+2i\lambda u_x-u_{xx}-2u^3\\
\noalign{\medskip}
-4\lambda^2u+2i\lambda u_x+u_{xx}+2u^3&
4i\lambda^3-2i\lambda u^2\endbmatrix,$$
and the compatibility condition (5.2) yields
the mKdV equation (2.5).

As for the first-order system
$v_x=\Cal X v,$ where
$$\Cal X:=\bmatrix
-i\lambda&-\displaystyle\frac12\,u_x(x,t)\\
\noalign{\medskip}
\displaystyle\frac12\,u_x(x,t)&i\lambda\endbmatrix,$$
we obtain the matrix operator $\Cal T$ as
$$\Cal T=\displaystyle\frac{i}{4\lambda}\bmatrix
\cos u&\sin u\\
\noalign{\medskip}
\sin u&
-\cos u\endbmatrix.$$
Then, the compatibility condition (5.2) gives us
the sine-Gordon equation
$$u_{xt}=\sin u.$$

\vskip 15 pt
\noindent {{\bf {\larger VI. Direct Scattering Problem}}}
\vskip 3 pt

The direct scattering problem consists of determining the scattering data
when the potential is known. This problem is usually solved by obtaining
certain specific solutions, known as the Jost solutions, to the relevant LODE.
The appropriate scattering data can be constructed with the help of
spatial asymptotics of the Jost solutions at infinity or from certain
Wronskian relations among the Jost solutions. In this section we
review the scattering data corresponding to the Schr\"odinger equation
(2.2) and to the Zakharov-Shabat system (2.4).
The scattering data sets for other LODEs can similarly be obtained.

Consider (2.2) at fixed $t$ by assuming that
the potential $u(x,t)$ belongs to the Faddeev class, i.e. $u(x,t)$
is real valued and
$\int_{-\infty}^\infty dx\,(1+|x|)\,|u(x,t)|$ is finite.
The Schr\"odinger equation has two types of solutions;
namely, scattering solutions
and bound-state solutions. The scattering solutions are those
that consist of
linear combinations of $e^{ikx}$ and $e^{-ikx}$ as
$x\to\pm\infty,$ and they occur for
$k\in\bold R\setminus\{0\},$
i.e. for real nonzero values of $k.$
Two linearly independent scattering solutions $f_{\text l}$
and $f_{\text r},$ known as the
Jost solution from the left and from the right, respectively,
are those solutions to (2.2) satisfying
the respective asymptotic conditions
$$f_{\text l}(k,x,t)=e^{ikx}+o(1),\quad
f'_{\text l}(k,x,t)=ike^{ikx}+o(1),\qquad
x\to+\infty,\tag 6.1$$
$$f_{\text r}(k,x,t)=e^{-ikx}+o(1),\quad
f'_{\text r}(k,x,t)=-ike^{-ikx}+o(1),\qquad
x\to-\infty,$$
where the notation $o(1)$ indicates the quantities
that vanish.
Writing their remaining spatial asymptotics in the form
$$f_{\text l}(k,x,t)=\displaystyle\frac{e^{ikx}}{T(k,t)}
+\displaystyle\frac{L(k,t)\,e^{-ikx}}{T(k,t)}+o(1),
\qquad x\to-\infty,\tag 6.2$$
$$f_{\text r}(k,x,t)=\displaystyle\frac{e^{-ikx}}{T(k,t)}
+\displaystyle\frac{R(k,t)\,e^{ikx}}{T(k,t)}+o(1),
\qquad x\to+\infty,$$
we obtain the scattering coefficients;
namely, the transmission coefficient $T$
and the reflection coefficients
$L$ and $R,$ from the left and right, respectively.

Let $\bold C^+$ denote the upper half complex plane.
A bound-state solution to
(2.2) is a solution that belongs to
$L^2(\bold R)$ in the $x$ variable. Note that
$L^2(\bold R)$ denotes the set of
complex-valued functions whose absolute squares
are integrable on the real line $\bold R.$
When $u(x,t)$ is in the Faddeev class, it is known
[5,7-9,16-19] that
the number of bound states is finite,
the multiplicity of each bound state is one,
and the bound-state solutions can occur only at certain $k$-values
on the imaginary axis in $\bold C^+.$
Let us use $N$ to denote
the number of bound states,
and suppose that the bound states occur at $k=i\kappa_j$ with
the ordering $0<\kappa_1<\dots<\kappa_N.$
Each bound state corresponds to a pole of
$T$ in $\bold C^+.$ Any bound-state solution at $k=i\kappa_j$
is a constant multiple of
$f_{\text l}(i\kappa_j,x,t).$
The left and right bound-state norming constants $c_{{\text l}j}(t)$
and $c_{{\text r}j}(t),$ respectively, can be defined as
$$c_{{\text l}j}(t):=\left[\displaystyle\int_{-\infty}
^\infty dx\, f_{\text l}(i\kappa_j,x,t)^2\right]^{-1/2},\quad
c_{{\text r}j}(t):=\left[\displaystyle\int_{-\infty}
^\infty dx\, f_{\text r}(i\kappa_j,x,t)^2\right]^{-1/2},$$
and they are related to each other through the residues of $T$ via
$$\text{Res}\,(T,i\kappa_j)=i\,c_{{\text l}j}(t)^2\,\gamma_j(t)
=i\,\displaystyle\frac{c_{{\text r}j}(t)^2}{\gamma_j(t)},\tag 6.3$$
where the $\gamma_j(t)$ are the dependency constants defined as
$$\gamma_j(t):=\displaystyle\frac{f_{\text l}
(i\kappa_j,x,t)}{f_{\text r}(i\kappa_j,x,t)}.
\tag 6.4$$
The sign of $\gamma_j(t)$ is the same as that of
$(-1)^{N-j},$ and hence
$c_{{\text r}j}(t)=(-1)^{N-j}\gamma_j(t)\, c_{{\text l}j}(t).$

The scattering matrix associated with (2.2)
consists of the transmission coefficient $T$
and the two reflection coefficients $R$ and $L,$
and it can be constructed
 from $\{\kappa_j\}_{j=1}^N$ and one of the
reflection coefficients.
For example, if we start with the right reflection
coefficient $R(k,t)$ for $k\in\bold R,$ we get
$$T(k,t)=\left(\displaystyle\prod_{j=1}^N\displaystyle
\frac{k+i\kappa_j}{k-i\kappa_j}\right)
\,\exp\left(\displaystyle\frac{1}{2\pi i}\int_{-\infty}^\infty
ds\,\displaystyle\frac{\log(1-|R(s,t)|^2)}
{s-k-i0^+}\right), \qquad
k\in\bold C^+\cup\bold R,$$
where the quantity $i0^+$ indicates that the value
for $k\in\bold R$ must be obtained
as a limit from $\bold C^+.$
Then, the left reflection coefficient $L(k,t)$ can be constructed via
$$L(k,t)=-\displaystyle\frac{\overline{R(k,t)}
\,T(k,t)}{\overline{T(k,t)}},\qquad k\in\bold R.$$
We will see in the next section that $T(k,t)=T(k,0),$
$|R(k,t)|=|R(k,0)|,$ and $|L(k,t)|=|L(k,0)|.$

For a detailed study of the direct scattering problem for the
1-D Schr\"odinger equation, we refer the reader to
[5,7-9,16-19]. It is important to remember that
$u(x,t)$ for $x\in\bold R$ at each fixed $t$ is uniquely
determined [5,7-9,16-18] by the scattering data
$\big\{R,\{\kappa_j\},\{c_{{\text l}j}(t)\}\big\}$
or one of its equivalents.
Letting $c_j(t):=c_{{\text l}j}(t)^2,$ we will work with
one such data set, namely $\big\{R,\{\kappa_j\},\{c_j(t)\}\big\},$
in Sections VII and VIII.

Having described the scattering data associated with the
Schr\"odinger equation, let us briefly describe the scattering data
associated with the Zakharov-Shabat system (2.4). Assuming that
$u(x,t)$ for each $t$ is integrable in $x$ on $\bold R,$
the two Jost solutions
$\psi(\lambda,x,t)$ and $\phi(\lambda,x,t),$ from the left and from the
right, respectively, are those unique solutions to (2.4)
satisfying the respective asymptotic conditions
$$\psi(\lambda,x,t)=\bmatrix 0\\
\noalign{\medskip}
e^{i\lambda x}\endbmatrix+o(1), \quad x\to +\infty;\qquad
\phi(\lambda,x,t)=\bmatrix e^{-i\lambda x}\\
\noalign{\medskip}
0\endbmatrix+o(1), \quad x\to -\infty.\tag 6.5$$
The transmission coefficient $T,$ the
left reflection coefficient $L,$ and the
right reflection coefficient $R$ are obtained via the
asymptotics
$$\psi(\lambda,x,t)=\bmatrix \displaystyle\frac{L(\lambda,t)\,e^{-i\lambda x}}{T(\lambda,t)}\\
\noalign{\medskip}
\displaystyle\frac{e^{i\lambda x}}{T(\lambda,t)}\endbmatrix+o(1), \quad x\to -\infty;\qquad
\phi(\lambda,x,t)=\bmatrix \displaystyle\frac{e^{-i\lambda x}}{T(\lambda,t)}\\
\noalign{\medskip}
\displaystyle\frac{R(\lambda,t)\,e^{i\lambda x}}{T(\lambda,t)}\endbmatrix+o(1),
\quad x\to +\infty.\tag 6.6$$
The bound-state solutions to (2.4) occur at those $\lambda$ values
corresponding to the poles of $T$ in $\bold C^+.$
Let us use
$\{\lambda_j\}_{j=1}^N$ to denote the set of such poles.
It should be noted that such poles are not necessarily located
on the positive imaginary axis. Furthermore, unlike
the Schr\"odinger equation, the multiplicities of
such poles may be greater than one. Let us assume that the
pole $\lambda_j$ has multiplicity
$n_j.$ Corresponding to the pole $\lambda_j,$ one
associates [4,20] $n_j$ bound-state
norming constants $c_{js}(t)$ for $s=0,1,\dots,n_j-1.$
We assume that, for each fixed $t,$ the potential
$u(x,t)$ in the Zakharov-Shabat system is uniquely
determined by the scattering data
$\big\{R,\{\lambda_j\},\{c_{js}(t)\}\big\}$ and vice versa.

\vskip 15 pt
\noindent {{\bf {\larger VII. Time Evolution of the Scattering Data}}}
\vskip 3 pt

As the initial profile $u(x,0)$ evolves to $u(x,t)$ while satisfying the
NPDE, the corresponding initial scattering data $S(\lambda,0)$ evolves
to $S(\lambda,t).$ Since the scattering data can be obtained from the Jost
solutions to the associated LODE, in order to determine the time evolution of
the scattering data, we can analyze the time evolution of the
Jost solutions with the help of the Lax method or the AKNS method.

Let us illustrate how to
determine the time evolution of the scattering data in the
Schr\"odinger equation with the help of the Lax method.
As indicated in Section IV, the spectral parameter $k$ and hence also
the values $\kappa_j$ related to the bound states remain
unchanged in time. Let us obtain the time evolution
of $f_{\text l}(k,x,t),$ the Jost solution from the left.
 From condition (ii) in Section IV, we see that
the quantity $\partial_t f_{\text l}-\Cal A
f_{\text l}$ remains a solution to
(2.2) and hence we can write it
as a linear combination of the two linearly
independent Jost solutions $f_{\text l}$ and $f_{\text r}$ as
$$\partial_t f_{\text l}(k,x,t)-\left(-4\partial_x^3
+6u\partial_x+3u_x\right)
f_{\text l}(k,x,t)=p(k,t)\,f_{\text l}(k,x,t)
+q(k,t)\,f_{\text r}(k,x,t),\tag 7.1$$
where the coefficients $p(k,t)$ and $q(k,t)$
are yet to be determined and $\Cal A$
is the operator in (4.7).
For each fixed $t,$ assuming
$u(x,t)=o(1)$ and $u_x(x,t)=o(1)$
as $x\to+\infty$ and
using (6.1) and (6.2) in (7.1) as $x\to+\infty,$ we get
$$\partial_t e^{ikx}+4\partial_x^3 e^{ikx}=p(k,t)\, e^{ikx}
+q(k,t)\left[\displaystyle\frac{1}{T(k,t)}\,e^{-ikx}
+\displaystyle\frac{R(k,t)}{T(k,t)}\,e^{ikx}\right]+o(1).\tag 7.2$$
Comparing the coefficients
of $e^{ikx}$ and $e^{-ikx}$ on the two sides of (7.2), we obtain
$$q(k,t)=0,\quad p(k,t)=-4ik^3.$$
Thus, $f_{\text l}(k,x,t)$ evolves in time
by obeying the linear
third-order PDE
$$\partial_t f_{\text l}-\Cal A f_{\text l}=
-4ik^3 f_{\text l}.\tag 7.3$$
Proceeding in a similar manner, we find that
$f_{\text r}(k,x,t)$ evolves in time according to
$$\partial_t f_{\text r}-\Cal A
f_{\text r}=4ik^3f_{\text r}.\tag 7.4$$

Notice that the time evolution of each Jost solution is
fairly complicated. We will see, however, that the
time evolution of the scattering data is very simple.
Letting $x\to -\infty$ in (7.3),
using (6.2) and
$u(x,t)=o(1)$ and $u_x(x,t)=o(1)$
as $x\to-\infty,$ and comparing the coefficients
of $e^{ikx}$ and $e^{-ikx}$ on both sides, we obtain
$$\partial_t T(k,t)=0,\quad \partial_t L(k,t)=-8ik^3 L(k,t),$$
yielding
$$T(k,t)=T(k,0),\quad
L(k,t)=L(k,0)\,e^{-8ik^3t}.$$
In a similar way, from (7.4) as $x\to+\infty,$ we get
$$R(k,t)=R(k,0)\,e^{8ik^3t}
.\tag 7.5$$
Thus, the transmission coefficient remains unchanged
and only the phases of the reflection coefficients
change as time progresses.

Let us also evaluate the time evolution of
the dependency constants $\gamma_j(t)$
defined in (6.4).
Evaluating (7.3) at $k=i\kappa_j$
and replacing $f_{\text l}(i\kappa_j,x,t)$ by
$\gamma_j(t)f_{\text r}(i\kappa_j,x,t),$
we get
$$f_{\text r}(i\kappa_j,x,t)\,\partial_t\gamma_j(t)
+\gamma_j(t)\,\partial_t
f_{\text r}(i\kappa_j,x,t)-\gamma_j(t)
\Cal A f_{\text r}(i\kappa_j,x,t)=-4\kappa_j^3
\gamma_j(t)f_{\text r}(i\kappa_j,x,t).\tag 7.6$$
On the other hand, evaluating (7.4) at $k=i\kappa_j,$ we obtain
$$\gamma_j(t)\,\partial_t
f_{\text r}(i\kappa_j,x,t)-\gamma_j(t)\,
\Cal A f_{\text r}(i\kappa_j,x,t)=4\kappa_j^3
\gamma_j(t)\,f_{\text r}(i\kappa_j,x,t).\tag 7.7$$
Comparing (7.6) and (7.7) we see that
$\partial_t \gamma_j(t)=-8\kappa_j^3 \gamma_j(t),$ or equivalently
$$\gamma_j(t)=\gamma_j(0)\,e^{-8\kappa_j^3 t}
.\tag 7.8$$
Then, with the help of (6.3) and (7.8), we determine the
time evolutions of
the norming constants as
$$c_{{\text l}j}(t)=c_{{\text l}j}(0)\,e^{4\kappa_j^3 t}
,\quad
c_{{\text r}j}(t)=c_{{\text r}j}(0)\,e^{-4\kappa_j^3 t}
.$$
The norming constants $c_j(t)$ appearing in the Marchenko kernel
(8.1) are related to $c_{{\text l}j}(t)$ as
$c_j(t):=c_{{\text l}j}(t)^2,$ and hence their time evolution is described
as
$$c_j(t)=c_j(0)\,e^{8\kappa_j^3 t}.\tag 7.9$$

As for the NLS equation and other integrable NPDEs, the time
evolution of the related scattering data sets can be obtained in a similar
way. For the former, in terms of the operator $\Cal A$ in (4.8), the Jost solutions
$\psi(\lambda,x,t)$ and $\phi(\lambda,x,t)$ appearing
in (6.5) evolve according to the respective linear PDEs
$$\psi_t-\Cal A \psi=-2i\lambda^2\psi,\quad
\phi_t-\Cal A \phi=2i\lambda^2\phi.$$
The scattering coefficients appearing in (6.6) evolve
according to
$$T(\lambda,t)=T(\lambda,0), \quad
R(\lambda,t)=R(\lambda,0)\,e^{4i\lambda^2 t},\quad
L(\lambda,t)=L(\lambda,0)\,e^{-4i\lambda^2 t}.\tag 7.10$$
Associated with the bound-state pole $\lambda_j$ of $T,$ we have
the bound-state norming constants $c_{js}(t)$ appearing in the
Marchenko kernel $\Omega(y,t)$ given in (8.4). Their time
evolution is governed [4] by
$$\bmatrix c_{j(n_j-1)}(t)&
c_{j(n_j-2)}(t)&\dots
& c_{j 0}(t)\endbmatrix= \bmatrix c_{j(n_j-1)}(0)&
c_{j(n_j-2)}(0)& \dots & c_{j
0}(0)\endbmatrix e^{-4i A_j^2t},\tag 7.11$$
where the $n_j\times n_j$ matrix $A_j$ appearing in the exponent is defined as
$$A_j:=\bmatrix -i\lambda_j&
-1&0&\dots&0&0\\
0& -i\lambda_j& -1&\dots&0&0\\
0&0&-i\lambda_j& \dots&0&0\\
\vdots&\vdots &\vdots &\ddots&\vdots&\vdots\\
0&0&0&\dots&-i\lambda_j&-1\\
0&0&0&\dots&0&-i\lambda_j\endbmatrix.$$

\vskip 15 pt
\noindent {{\bf {\larger VIII. Inverse Scattering Problem}}}
\vskip 3 pt

In Section VI we have seen how the initial scattering data $S(\lambda,0)$
can be constructed from the initial profile $u(x,0)$ of the potential
by solving the direct scattering problem for the
relevant LODE.
Then, in Section VII we have seen how to obtain the time-evolved
scattering data $S(\lambda,t)$ from the initial scattering data
$S(\lambda,0).$ As the final step in the IST, in this section we outline
how to obtain $u(x,t)$ from $S(\lambda,t)$
by solving the relevant inverse scattering problem.
Such an inverse scattering problem may be solved by
the Marchenko method [5,7-9,16-19]. Unfortunately, in the literature
many researchers refer to this method as the Gel'fand-Levitan
method or the Gel'fand-Levitan-Marchenko method, both of which are misnomers.
The Gel'fand-Levitan method [5,7,16,17,19] is a different method to solve
the inverse scattering problem, and the corresponding Gel'fand-Levitan integral
equation involves an integration on the finite
interval $(0,x)$ and its kernel
is related to the Fourier transform of the spectral measure associated with
the LODE. On the other hand, the Marchenko integral
equation involves an integration on the semi-infinite
interval $(x,+\infty),$ and its kernel
is related to the Fourier transform of the
scattering data.

In this section we first outline the recovery of the solution $u(x,t)$
to the KdV equation from the corresponding
time-evolved scattering data $\big\{R,\{\kappa_j\},\{c_j(t)\}\big\}$
appearing in (7.5) and (7.9). Later, we will also outline the recovery of
the solution $u(x,t)$
to the NLS equation from the corresponding
time-evolved scattering data $\big\{R,\{\lambda_j\},\{c_{js}(t)\}\big\}$
appearing in (7.10) and (7.11).

The solution $u(x,t)$
to the KdV equation (2.1)
can be obtained from the time-evolved scattering data
by using the Marchenko method as follows:

\item{(a)} From the scattering data $\big\{R(k,t),\{\kappa_j\},\{c_j(t)\}\big\}$
appearing in (7.5) and (7.9), form the
Marchenko kernel $\Omega$ defined via
$$\Omega(y,t):=\displaystyle
\frac{1}{2\pi}\int_{-\infty}^\infty
dk\,R(k,t)\,e^{iky}+\sum_{j=1}^N
c_{j}(t)\,e^{-\kappa_jy}.\tag 8.1$$

\item{(b)} Solve the corresponding Marchenko integral equation
$$K(x,y,t)+\Omega(x+y,t)+\int_x^\infty dz
\,K(x,z,t)\,\Omega(z+y,t)=0,\qquad
x<y<+\infty,\tag 8.2$$
and obtain its solution $K(x,y,t).$

\item{(c)} Recover $u(x,t)$
by using
$$u(x,t)=-2\,\displaystyle\frac{\partial K(x,x,t)}{
\partial x}.\tag 8.3$$

The solution $u(x,t)$ to the
NLS equation (2.3) can be obtained from the time-evolved scattering data
by using the Marchenko method as follows:

\item{(i)} From the scattering data $\big\{R(\lambda,t),\{\lambda_j\},
\{c_{js}(t)\}\big\}$ appearing in (7.10) and (7.11), form the
Marchenko kernel $\Omega$ as
$$\Omega(y,t):=\displaystyle\frac{1}{2\pi}\displaystyle\int_{-\infty}^\infty d\lambda\,
R(\lambda,t)\,e^{i\lambda y}+\sum_{j=1}^{N}\sum_{s=0}^{n_j-1} c_{js}(t)
\displaystyle\frac{y^s}{s!}\,e^{i\lambda_j y}.
\tag 8.4$$

\item{(ii)} Solve the Marchenko integral equation
$$K(x,y,t)-\overline{\Omega(x+y,t)}+\int_x^\infty dz\int_x^\infty ds\, K(x,s,t)\,
\Omega(s+z,t)\,\overline{\Omega(z+y,t)}=0,\qquad
x<y<+\infty,$$
and obtain its solution $K(x,y,t).$

\item{(iii)} Recover $u(x,t)$ from the solution $K(x,y,t)$ to the
Marchenko equation via
$$u(x,t)=-2 K(x,x,t).$$

\item{(iv)} Having determined $K(x,y,t),$ one can alternatively get
$|u(x,t)|^2$ from
$$|u(x,t)|^2=2\,\displaystyle\frac{\partial G(x,x,t)}{\partial x},$$
where we have defined
$$G(x,y,t):=-\int_x^\infty dz\, \overline{K(x,z,t)}\,\overline{\Omega(z+y,t)}.$$

\vskip 15 pt
\noindent {{\bf {\larger IX. Solitons}}}
\vskip 3 pt

A soliton solution to an
integrable NPDE is a solution $u(x,t)$ for which the reflection coefficient
in the corresponding scattering data is zero. In other words, a soliton solution
$u(x,t)$ to an integrable NPDE is nothing but a reflectionless potential in the associated LODE.
When the reflection coefficient is zero,
the kernel of the relevant Marchenko integral equation
becomes separable. An integral equation with a
separable kernel can be solved explicitly by transforming that linear
equation into a system of linear algebraic equations.
In that case, we get exact solutions to the integrable NPDE, which are
known as soliton solutions.

For the KdV equation the $N$-soliton solution is obtained by using
$R(k,t)=0$ in (8.1). In that case, letting
$$X(x):=\bmatrix e^{-\kappa_1 x}&e^{-\kappa_2 x}&\dots &e^{-\kappa_N x}\endbmatrix,
\quad Y(y,t):=\bmatrix c_1(t)\,e^{-\kappa_1 y}\\
\noalign{\medskip}
c_2(t)\,e^{-\kappa_2 y}\\
\vdots\\
c_N(t)\,e^{-\kappa_N y}\endbmatrix,$$
we get $\Omega(x+y,t)=X(x)\,Y(y,t).$ As a result of this separability
the Marchenko integral equation
can be solved algebraically and the solution
has the form
$K(x,y,t)=H(x,t)\,Y(y,t),$ where $H(x,t)$ is a row vector with
$N$ entries that are functions of $x$ and $t.$ A substitution in (8.2) yields
$$K(x,y,t)=-X(x)\,\Gamma(x,t)^{-1}Y(y,t),\tag 9.1$$
where the $N\times N$ matrix $\Gamma(x,t)$ is given by
$$\Gamma(x,t):=I+\int_x^\infty dz \,Y(z,t)\,X(z),\tag 9.2$$
with $I$ denoting the $N\times N$ identity
matrix.
Equivalently, the $(j,l)$-entry of $\Gamma$ is given by
$$\Gamma_{jl}=\delta_{jl}+\displaystyle\frac{c_j(0)\,
e^{-2 \kappa_j x+8\kappa_j^3 t}}{
\kappa_j +\kappa_l},$$
with $\delta_{jl}$ denoting the Kronecker delta.
Using (9.1) in (8.3) we obtain
$$u(x,t)=2\,\displaystyle\frac{\partial}{\partial x}\left[
X(x)\,\Gamma(x,t)^{-1}Y(x,t)\right]=
2\,\text{tr}\,\displaystyle\frac{\partial}{\partial x}\left[
Y(x,t)\,X(x)\,\Gamma(x,t)^{-1}\right],$$
where tr denotes the matrix trace (the sum of diagonal entries
in a square matrix). From
(9.2) we see that $-Y(x,t)\,X(x)$ is equal to the
$x$-derivative of $\Gamma(x,t)$ and hence the $N$-soliton solution can
also be written as
$$u(x,t)=
-2\,\text{tr}\,\displaystyle\frac{\partial}{\partial x}\left[
\displaystyle\frac{\partial \Gamma(x,t)}{\partial x}\,
\Gamma(x,t)^{-1}\right]=
-2\,\displaystyle\frac{\partial}{\partial x}\left[\displaystyle\frac{\frac{\partial}{\partial x}
\det \Gamma(x,t)}{\det \Gamma(x,t)}\right],\tag 9.3$$
where det denotes the matrix determinant. When
$N=1,$ we can express the one-soliton solution $u(x,t)$
to the KdV equation in the equivalent form
$$u(x,t)=-2\,\kappa_1^2\, \text{sech}^2\left(\kappa_1 x-4\kappa_1^3 t+\theta\right),$$
with $\theta:=\log\sqrt{2\kappa_1/c_1(0)}.$

Let us mention that, using
matrix exponentials, we can express [6]
the $N$-soliton solution appearing in (9.3) in various other
equivalent forms such as
$$u(x,t)=-4Ce^{-Ax+8A^3t}\Gamma(x,t)^{-1}
A\Gamma(x,t)^{-1}e^{-Ax}B,$$
where
$$A:=\text{diag}\{\kappa_1,\kappa_2,\dots,\kappa_N\},$$
$$B^\dagger:=\bmatrix 1&1&\dots&
1\endbmatrix,\quad
C:=\bmatrix c_1(0)&c_2(0)&\dots&c_N(0)\endbmatrix.\tag 9.4$$
Note that
a dagger is used for the matrix adjoint (transpose and complex conjugate),
and $B$ has $N$ entries. In this notation we can express (9.2) as
$$\Gamma(x,t)=I+\int_x^\infty dz \,e^{-zA}BC e^{-zA}
e^{8tA^3}.$$

As for the NLS equation, the well-known $N$-soliton solution (with simple
bound-state poles) is obtained by
choosing $R(\lambda,t)=0$ and $n_j=1$ in (8.4). Proceeding
as in the KdV case, we obtain the $N$-soliton solution in terms of
the triplet $A,$ $B,$ $C$ with
$$A:=\text{diag}\{-i\lambda_1,-i\lambda_2,\dots,-i\lambda_N\},\tag 9.5$$
where the complex constants $\lambda_j$ are the
distinct poles of the transmission coefficient in $\bold C^+,$
$B$ and $C$ are as in (9.4) except for the fact that the constants
$c_j(0)$ are now allowed to be nonzero complex numbers.
In terms of the matrices $P(x,t),$ $M,$ and $Q$ defined as
$$P(x,t):=\text{diag}\{e^{2i\lambda_1x+4i\lambda_1^2 t},
e^{2i\lambda_2 x+4i\lambda_2^2 t},
\dots,e^{2i\lambda_N x+4i\lambda_N^2 t}\},\quad
M_{jl}:=\displaystyle\frac{i}{\lambda_j-\overline\lambda_l},\quad
Q_{jl}:=\displaystyle\frac{-i\overline c_j c_l}{\overline\lambda_j-\lambda_l}.$$
we construct the $N$-soliton solution $u(x,t)$ to the NLS equation as
$$u(x,t)=-2B^\dagger \left[I+P(x,t)^\dagger Q P(x,t)\,M\right]^{-1} P(x,t)^\dagger
C^\dagger,\tag 9.6$$
or equivalently as
$$u(x,t)=-2B^\dagger e^{-A^\dagger x}\Gamma(x,t)^{-1}
e^{-A^\dagger x+4i(A^\dagger)^2t}C^\dagger,\tag 9.7$$
where we have defined
$$\Gamma(x,t):=I+\left[
\int_x^\infty ds\,\left(Ce^{-As-4iA^2t}\right)^\dagger
\left(Ce^{-As-4iA^2t}\right)\right]\left[
\int_x^\infty dz\,
\left(e^{-A z}B\right) \left(e^{-A z}B\right)^\dagger\right]
.\tag 9.8$$
Using (9.4) and (9.5) in (9.8), we get the $(j,l)$-entry of $\Gamma(x,t)$
as
$$\Gamma_{jl}=
\delta_{jl}-\displaystyle
\sum_{m=1}^N\displaystyle\frac {\overline{c}_j c_l\,
e^{i(2\lambda_m-\overline{\lambda}_j-\overline{\lambda}_l)x
+4i(\lambda_m^2-\overline{\lambda}_j^2)t}}
{(\lambda_m-\overline{\lambda}_j)
(\lambda_m-\overline{\lambda}_l)}.$$
Note that the absolute square of $u(x,t)$ is given by
$$|u(x,t)|^2=\text{tr}\left[\displaystyle\frac
{\partial}{\partial x}
\left(\Gamma(x,t)^{-1}\displaystyle\frac{\partial \Gamma(x,t)}{\partial x}
\right)\right]=
\displaystyle\frac{\partial}{\partial x}\left[\displaystyle\frac{\frac{\partial}{\partial x}
\det \Gamma(x,t)}{\det \Gamma(x,t)}\right].$$
For the NLS equation, when $N=1,$ from (9.6) or (9.7)
we obtain the single-soliton solution
$$u(x,t)=\displaystyle\frac{-8 \overline c_1 (\text{Im}[\lambda_1])^2 \,e^{-2i \overline\lambda_1
x-4i (\overline\lambda_1)^2t}}
{4 (\text{Im}[\lambda_1])^2+|c_1|^2\,e^{-4x(\text{Im}[\lambda_1])-8t
(\text{Im}[\lambda_1^2])}},$$
where Im denotes the imaginary part.

\vskip 15 pt
\noindent {{\bf {\larger X. Future Directions}}}
\vskip 3 pt

There are many issues related to the IST and solitons that cannot be
discussed in such a short review. We will briefly mention only a few.

Can we characterize integrable NPDEs? In other words,
can we find a set of necessary and sufficient
conditions that guarantee that an IVP
for a NPDE is solvable via an IST?
Integrable
NPDEs seem
to have some common characteristic features [1] such as possessing
Lax pairs, AKNS pairs, soliton solutions, infinite
number of conserved quantities, a Hamiltonian formalism,
the Painlev\'e property, and the B\"acklund transformation.
Yet, there does not seem to be a satisfactory
solution to their characterization problem.

Another interesting question is the determination of the
LODE associated with an IST. In other words, given an
integrable NPDE, can we determine the corresponding
LODE? There does not yet seem to be a
completely satisfactory answer to this question.

When the initial scattering coefficients are rational functions
of the spectral parameter, representing the
time-evolved scattering data in terms of matrix exponentials results
in the separability of the kernel of the Marchenko integral equation.
In that case, one obtains explicit formulas [4,6] for exact solutions to
some integrable NPDEs and such solutions are constructed
in terms of a triplet of constant matrices $A,$ $B,$ $C$
whose sizes are $p\times p,$ $p\times 1,$ and
$1\times p,$ respectively, for any positive integer $p.$
Some special cases of such solutions have
been mentioned in Section IX, and it would
be interesting to determine if such exact solutions can be constructed
also when $p$ becomes infinite.

\vskip 15 pt
\noindent {{\bf {\larger XI. Bibliography}}}
\vskip 3 pt

\vskip 10 pt
\noindent {{\bf {\larger Primary Literature}}
\vskip 3 pt

\item{[1]} M. J. Ablowitz and P. A. Clarkson,
{\it Solitons, nonlinear evolution equations and inverse scattering,}
Cambridge University Press, Cambridge, 1991.

\item{[2]} M. J. Ablowitz, D. J. Kaup, A. C. Newell, and
H. Segur, {\it Method for solving the sine-Gordon equation,}
Phys. Rev. Lett. {\bf 30}, 1262--1264 (1973).

\item{[3]} M. J. Ablowitz, D. J. Kaup, A. C. Newell, and H. Segur, {\it The
inverse scattering transform-Fourier analysis for nonlinear problems,} Stud.
Appl. Math. {\bf 53}, 249--315 (1974).

\item{[4]} T. Aktosun, F. Demontis, and C. van der Mee,
{\it Exact solutions to the focusing nonlinear Schr\"odinger equation,}
Inverse Problems {\bf 23}, 2171--2195 (2007).

\item{[5]} T. Aktosun and M. Klaus, {\it
Chapter 2.2.4,
Inverse theory: problem on the line,}
In: E. R. Pike and P. C. Sabatier (eds.),
{\it Scattering,} Academic Press, London, 2001,
pp. 770--785.

\item{[6]} T. Aktosun and C. van der Mee, {\it Explicit solutions to the
Korteweg-de Vries equation on the half-line,} Inverse Problems {\bf 22},
2165--2174 (2006).

\item{[7]} K. Chadan and P. C. Sabatier, {\it Inverse problems in quantum
scattering theory,} 2nd ed., Springer, New York, 1989.

\item{[8]} P. Deift and E. Trubowitz, {\it
Inverse scattering on the line,}
Commun. Pure Appl. Math. {\bf 32}, 121--251 (1979).

\item{[9]}
L. D. Faddeev,
{\it Properties of the $S$-matrix of the
one-dimensional Schr\"odinger equation,} Amer.
Math. Soc. Transl. (Ser.
2) {\bf 65}, 139--166 (1967).

\item{[10]} E. Fermi, {\it Collected papers,
Vol. II: United States, 1939--1954}, University of
Chicago Press, Chicago, 1965.

\item{[11]} E. Fermi, J. Pasta, and S. Ulam,
{\it Studies of non linear problems, I,} Document
LA-1940, Los Alamos National Laboratory, May 1955.

\item{[12]} C. S. Gardner, J. M. Greene, M. D. Kruskal and R. M. Miura,
{\it Method for solving the Korteweg-de Vries equation,}
Phys. Rev. Lett. {\bf 19}, 1095--1097 (1967).

\item{[13]} I. M. Gel'fand and B. M. Levitan,
{\it On the determination of a differential equation from its
spectral function,}
Amer. Math. Soc. Transl. (Ser. 2) {\bf 1}, 253--304 (1955).

\item{[14]} D. J. Korteweg and G. de Vries,
{\it On the change of form
of long waves advancing in a rectangular channel and on a new type of
long stationary waves,}
Phil. Mag. {\bf 39}, 422--443 (1895).

\item{[15]} P. D. Lax, {\it Integrals of nonlinear
equations of evolution and solitary waves,}
Commun. Pure Appl. Math. {\bf 21}, 467--490 (1968).

\item{[16]} B. M. Levitan, {\it
Inverse Sturm-Liouville problems,} VNU Science Press, Utrecht, 1987.

\item{[17]} V. A. Marchenko, {\it Sturm-Liouville operators and
applications,} Birk\-h\"au\-ser, Basel, 1986.

\item{[18]} A. Melin,
{\it Operator methods for inverse scattering
on the real line,} Commun. Partial Differential
Equations {\bf 10}, 677--766 (1985).

\item{[19]} R. G. Newton,
{\it The Marchenko and Gel'fand-Levitan methods in the inverse scattering
problem in one and three dimensions,} In: J. B. Bednar, R. Redner, E. Robinson,
and A. Weglein (eds.),
{\it Conference on inverse scattering: theory and
application,}
SIAM, Philadelphia, 1983, pp. 1--74.

\item{[20]} E. Olmedilla, {\it Multiple pole solutions of the nonlinear
Schr\"odinger equation,} Phys. D {\bf 25}, 330--346 (1987).

\item{[21]} J. S. Russell, {\it Report on waves,} Report of
the 14th meeting of the
British Association for the Advancement of  Science,
John Murray, London, 1845, pp. 311--390.

\item{[22]} M. Wadati, {\it The exact solution of the
modified Korteweg-de Vries equation,}
J. Phys. Soc. Japan {\bf 32}, 1681 (1972).

\item{[23]} N. J. Zabusky and M. D. Kruskal,
{\it Interaction of ``solitons" in a collisionless plasma and
the recurrence of initial states,} Phys. Rev. Lett. {\bf 15}, 240--243
(1965).

\item{[24]} V. E. Zakharov
and A. B. Shabat,
{\it Exact theory of two-dimensional self-focusing and
one-dimensional self-modulation of waves
in nonlinear media,} Soviet Phys. JETP {\bf 34},
62--69 (1972).

\item{[25]} http://www.osti.gov/accomplishments/pdf/A80037041/A80037041.pdf

\vskip 10 pt
\noindent {{\bf {\larger Books and Reviews}}
\vskip 3 pt

\item{} M. J. Ablowitz and H. Segur, {\it
Solitons and the inverse scattering
transform,} SIAM, Philadelphia, 1981.

\item{} T. Aktosun,
{\it Inverse scattering transform, KdV, and solitons,}
In: J. A. Ball, J. W. Helton, M. Klaus, and L. Rodman (eds.),
{\it Current trends in operator theory and its applications,}
Birkh\"auser, Basel, 2004, pp. 1--22.

\item{} T. Aktosun,
{\it Solitons and inverse scattering transform,}
In: D. P. Clemence and G. Tang (eds.),
{\it Mathematical studies in nonlinear wave propagation,}
Contemporary Mathematics, Vol. 379,
Amer. Math. Soc., Providence, 2005, pp. 47--62.

\item{} R. K. Dodd, J. C. Eilbeck, J. D. Gibbon, and H. C. Morris,
{\it Solitons and nonlinear wave equations,} Academic Press, London, 1982.

\item{} P. G. Drazin and R. S. Johnson, {\it Solitons: an introduction,}
Cambridge University Press, Cambridge, 1988.

\item{} G. L. Lamb, Jr., {\it Elements of soliton theory,}
Wiley, New York, 1980.

\item{} S. Novikov, S. V. Manakov, L. P. Pitaevskii, and V. E. Zakharov,
{\it Theory of solitons,}
Consultants Bureau, New York, 1984.

\item{} A. C. Scott, F. Y. F. Chu and D. McLaughlin,
{\it The soliton: a new concept in applied science,} Proc. IEEE {\bf 61},
1443--1483 (1973).

\end